\documentclass{IEEEtaes}

\usepackage{color,array,amsthm}
\usepackage{graphicx}

\jvol{}
\jnum{}
\jmonth{}
\paper{}
\pubyear{2024}
\doiinfo{}

\usepackage[caption=false, font=footnotesize]{subfig}
\usepackage{rotating}
\usepackage{pdflscape}
\usepackage{booktabs}
\usepackage{siunitx}
\usepackage{mathtools}
\usepackage{cuted}
\usepackage{multirow}
\usepackage{eqnarray}
\usepackage{amsmath}
\usepackage{placeins}
\usepackage{float}
\usepackage{lipsum}
\usepackage{flushend}
\usepackage{balance}
\usepackage{scalerel}
\usepackage{stfloats}
\usepackage{bigints}
\usepackage{textcomp}
\usepackage{tabularx}
\usepackage{cleveref}

\usepackage{amsmath,amssymb,amsfonts,cite,graphicx,xcolor,color,multirow,array,eqnarray}
\usepackage{kantlipsum,lipsum,mathtools,tabularx,threeparttable,adjustbox,pdflscape,longtable,bigints}
\usepackage{flushend,booktabs,siunitx,textcomp,placeins,scalerel,cuted,breqn,csquotes,mwe,rotating,url}

\begin{document}

\title{Continuous Power Beaming to Lunar Far Side from EMLP-2 Halo Orbit}

\author{BARIS DONMEZ }
\member{Graduate Student Member, IEEE}
\affil{Polytechnique Montréal, Montréal, QC, Canada} 

\author{GUNES KARABULUT KURT}
\member{Senior Member, IEEE}
\affil{Polytechnique Montréal, Montréal, QC, Canada} 

\authoraddress{Baris Donmez and Gunes Karabulut Kurt are with the Department of Electrical Engineering and Poly-Grames Research Center, Polytechnique Montréal, Montréal, QC H3T 1J4, Canada, (e-mail: \href{mailto:baris.donmez@polymtl.ca}{baris.donmez@polymtl.ca}). 
}

\markboth{DONMEZ ET AL.}{Continuous Power Beaming to Lunar Far Side from EMLP~2 Halo Orbit}
\maketitle

\begin{abstract}This paper focuses on FSO-based wireless power transmission (WPT) from Earth-Moon Lagrangian Point-2 (EMLP-2) to a receiver optical antenna equipped with solar cells that can be located anywhere on the lunar far side (LFS). Different solar-powered satellite (SPS) configurations which are EMLP-2 located single stable satellite and EMLP-2 halo orbit revolving single, double, and triple satellites are evaluated in terms of 100\% LFS surface coverage percentage (SCP) and continuous Earth visibility. It is found that an equidistant triple satellite scheme on EMLP-2 halo orbit with a semi-major axis length of 15,000 km provides full SCP for LFS and it is essential for the continuous LFS wireless power transmission. In our proposed dynamic cislunar space model, geometric and temporal parameters of the Earth-Moon systems are used in affine transformations. Our dynamic model enables us to determine the full coverage time rate of a specific region such as the LFS southern pole. The outcomes show that the equidistant double satellite scheme provides SCP=100\% during 88.60\% time of these satellites' single revolution around the EMLP-2 halo orbit. Finally, the probability density function (PDF) of the random harvested power $P_H$ is determined and it validates the simulation data extracted from the stable EMLP-2 satellite and revolving satellite around EMLP-2 halo orbit for minimum and maximum LoS distances. Although the pointing devices to mitigate random misalignment errors are considered for the stable and revolving SPSs, better pointing accuracy is considered for the stable satellite. Our simulations show that the probability of $P_H\le$41.6~W is around 0.5 for the stable satellite whereas the CDF=0.99 for the revolving satellite case for a transmit power of 1 kW.   
\end{abstract}


\section{INTRODUCTION}
\label{Introduction}
T{\scshape he} Moon is the closest satellite to the Earth and hence it can be utilized as a base station for deep space exploration missions. For instance, a spacecraft can satisfy its needs (i.e., maintenance and battery recharge) before launching interplanetary missions. Moreover, asteroid and meteor impacts on the moon lead to the existence of platinum-group minerals on the lunar soil whereas solar winds enable Helium-3, which is very scarce on the Earth, to be found in the lunar regolith. Due to these rare and valuable resources on the Moon, lunar mining is considered vital since there is a finite available resource on Earth \cite{general1,lunargateway,lunarmining}. However, the tidal locking between the Moon and the Earth leads to a permanent unobservability of one side of the Moon (i.e., the lunar far side (LFS)) from the Earth. Since the Moon orbits around the world, there are 14 Earth days of total darkness, followed by the same amount of time for complete sunlight. This causes a mission interruption due to the run-out of batteries. Therefore, a solution for recharging the lunar equipment and vehicles while they are working on the LFS in complete darkness is necessary. Wireless power transfer (WPT) can be an option to overcome this problem. Since the considered distances are significantly large in point-to-point space link, and the losses that occur due to atmospheric attenuation, scintillation, and fiber-coupling are negligible \cite{ComprehensivePathLoss}, the free-space optics (FSO) communication technology can be considered as a proper candidate. 

In our previous work, an FSO-based inter-satellite energy harvesting system is modeled by considering realistic laser power and solar cell conversion efficiencies. A solar-powered satellite (SPS) beams power to a small satellite such as 1U (0.1$\times$0.1$\times$0.1 m) or 12U (0.2$\times$0.2$\times$0.3 m) with adaptive beam divergence to maintain spot diameters of 0.1 and 0.2 m, respectively. Independent random misalignment errors on both the transmitter and receiver sides are mitigated by proposing an acquisition, tracking, and pointing (ATP) system module \cite{donmezatp}.  
On the other hand, existing lunar orbits in the literature need to be assessed to find adequate cislunar orbits. The following characteristics of lunar orbits are critical in the decision-making process: the vicinity to the LFS, surface coverage percentage (SCP), and stability (i.e., station-keeping necessity). There are smaller and larger moon orbit options and they offer advantages and disadvantages so a trade-off must be made while making a decision.

Smaller moon orbit options are low lunar orbit (LLO), elliptical lunar orbit (ELO), prograde circular orbit (PCO), and frozen lunar orbit (FLO). The period of LLO is around 2 hours due to the altitude of 100 km thus LLO is favorable for lunar surface access (i.e., Apollo lunar exploration program). On the other hand, ELO, FLO, and PCO are elliptic orbits having substantial differences between perilune and apolune. The amplitude range of ELO, FLO, and PCO are 100--10,000, 880--8,800 km, and 3,000--5,000 km, respectively. Their orbital periods are similar and change between 11 to 14 hours. Moreover, these lunar orbits have different inclinations such that LLO can have any whereas ELO has only an equatorial inclination. Although these lunar orbits have different inclinations, they offer limited SCP since these are smaller lunar orbits. In general, smaller cislunar orbits compel satellites to maneuver frequently for station-keeping and these orbits cannot provide continuous LoS communication with the Earth due to the regular blackouts\cite{orbittypes}.

Larger moon orbit options are near rectilinear orbit (NRO), distant retrograde orbit (DRO), and Earth-Moon Lagrangian point (EMLP) halo orbits. Since these orbits are significantly large, the altitudes can reach even beyond 70,000 km and their period can be up to 14 days. NROs can be located over either the north or south pole of the moon and act as a bridge between EMLP L1 and L2 thus, they provide polar coverage. Unlike NROs, the DRO has a zero out-of-plane amplitude (i.e., Earth-Moon plane) thus, it provides lunar equator coverage only. It should be noted that a couple of hours of occultation occurs in every period of DRO. On the other hand, DRO and PCO provide foreseeable behaviors, and hence negligible corrective maneuvers are required \cite{orbittypes}. 

Five equilibrium, or libration, points are established due to the gravitational forces of moving two celestial bodies acting on each other and that is shown by the French mathematician Joseph Lagrange in 1772. Three of these points which are L1, L2, and L3 are collinear whereas L4 and L5 create equilateral triangles with these two rotating planets. In the case of Earth and Moon bodies, EMLPs L4 and L5 provide stability whereas L1, L2, and L3 require station-keeping. Since our focus is on the LFS, the halo orbit design around the EMLP L2 point which is around 65,000 km far from the lunar centroid, offers the best coverage among all and has a period of 8--14 days~\cite{orbittypes}. 
 
In the existing literature, FSO-based wireless power transfer in space is reviewed and future challenges are discussed \cite{zheng_wireless_2024}. The recent progress in the power conversion efficiencies (PCEs) of transmitting laser diodes and laser power converters (LPCs) for various wavelengths is summarized. 
Moreover, Farquhar focuses on finding a solution to the interrupted communication link between the Earth and the Lunar Far Side (LFS) by proposing a halo orbit on the EMLP-2 which is one of the libration points \cite{farquhar1971utilization}. The ratio between the semi-minor and semi-major axes to minimize the stationkeeping is computed as 0.343. The minimum semi-minor axis to pass over the lunar occultation and establish a continuous link with the Earth is computed as 3671 km \cite{farquhar1967}. 

Furthermore, \cite{Txaperturedia} focuses on the feasibility of WPT from EMLP-1 and -2 to a manned rover which requires 30 kW of harvested power. In addition, the required transmitter aperture diameter is computed as 50 m since the distance between these L1 and L2 points to the moon is beyond 60,000 km, and hence, a very narrow beam divergence angle is required. The laser wavelength of 800 nm and EHCE of 22.5 percent are preferred. In \cite{Nasa2000km3sat}, the feasibility of WPT to the lunar habitat and a rover is analyzed. Three satellites located 2,000 km away from the lunar surface form a triangle with equal angular separation. The laser wavelength of 850 nm and EHCE of 32 percent are considered in \cite{Nasa2000km3sat}. Besides, the required powers for the habitat and rover are 1 MW and 75 kW, respectively. In \cite{kerslake2008lunar}, laser power versus transmitter aperture diameter ( i.e., up to 3 m) and then versus solar cell receiver diameter (i.e., up to 5 m) are analyzed for 10 kW transferred power and 1 km lunar surface to surface WPT case.
In \cite{lopez2023lunar}, a polar LLO, an FLO, and a DRO in which three and ten satellite constellations are evaluated to provide 100 kW to the lunar base. The maximum laser power of 30 kW is considered for each solar-powered satellite.  

\begin{figure*}[t!]
\centering
\includegraphics[width=0.8\textwidth]{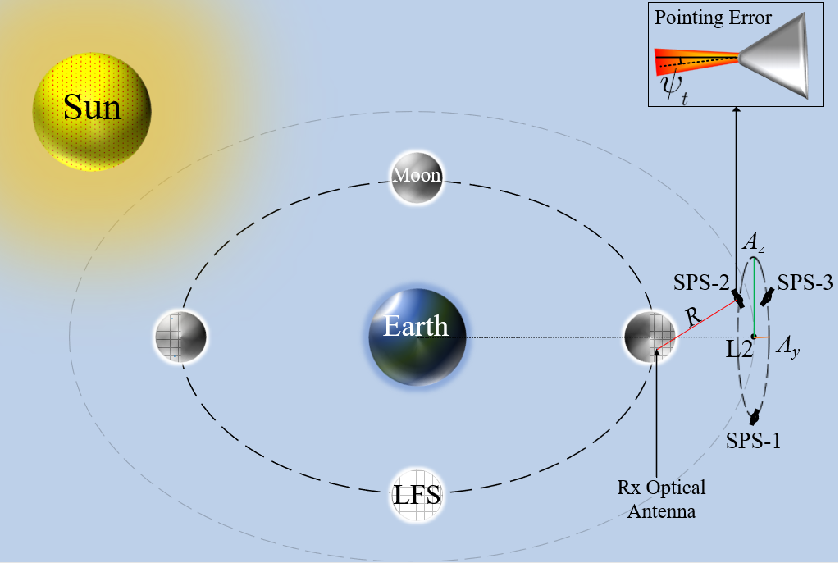}
\caption{System model of lunar far side optical power transmission and energy harvesting by a receiver equipped with solar cells.}
\label{fig1}
\vspace{-0.2cm}
\end{figure*}

There is a research gap on continuous energy harvesting on the lunar far side that enables a receiver (Rx) optical antenna equipped with solar cells to convert optical power to electrical power for the utilization of the research equipment during the 14 Earth days of eclipse. To fill this gap we select EMLP-2 and design a halo orbit with appropriate major and minor axis lengths to minimize the stationkeeping cost while providing continuous Earth visibility (i.e., deep space network stations (DSNSs)). We consider a dynamic cislunar space model with a minimum number of solar-powered satellites (SPS) that provide full LFS surface coverage so a receiver anywhere on LFS can be power-beamed continuously. In addition, the losses induced by the misalignment fading which are mitigated by using the beaconless pointing module (i.e., celestial references, internal sensors) \cite{lasertypes} due to the propagation time of 0.2 s are considered.      

Our paper investigates different EMLP-2 halo orbits and various SPS configurations that offer full LFS coverage and Earth visibility in our dynamic cislunar space model. The impact of the random misalignment error is considered in energy harvesting made at the receiver equipped with solar cells that can be located anywhere on LFS. The key contributions of this study can be listed as follows:
\vspace{-0.2cm}
\begin{itemize}
    \item We construct an EMLP-2 halo orbit by considering the semi-minor axis $A_y=0.343\times A_z$ \cite{farquhar1971utilization} and $A_y \geq 3671$ km \cite{farquhar1967} so there will be less stationkeeping cost, and continuous Earth visibility, respectively. 
     We analyze the surface coverages of the LFS for a single EMLP-2 stable satellite, a single EMLP-2 halo orbiting satellite, an EMLP-2 halo orbiting two satellites, and an EMLP-2 halo orbiting three satellites for various semi-major axis lengths $A_z$. 
  We conclude that equidistant three satellites revolving in EMLP-2 halo orbit with $A_z=15\times10^3$ km achieve continuous power beaming to a receiver optical antenna anywhere on the lunar far side due to SCP = 100\% and visibility to the Earth as well.
    \item We establish a time-based Earth-Moon-EMLP-2 halo orbit system presented in Fig. \ref{fig3} and it enables us to compute the full SCP duration of a specific region such as the promising south pole \cite{LunarSouthPole} of the LFS (Table \ref{table-LFS_SP_SCP_Duration}). Aside triple satellite scheme that provides continuous LFS full coverage, even a two-satellite configuration provides full coverage during 88.60\% of a full cycle around the EMLP-2 halo orbit.
     We show the contribution of each satellite to the overall SCP in detailed 3D figures at a sampled time for each scenario (Fig. \ref{fig7}). 
    \item We determine the probability density function (PDF) of the random harvested power and then compute its cumulative distribution function (CDF). It validates our simulation data extracted from the stable EMLP-2 satellite and a revolving halo orbit satellite whose pointing accuracies are different due to the different amounts of mechanical vibrations that occur on their transmitters (Figs. \ref{fig8}). The probability of $P_H\le$41.6~W is around 0.5 for the stable satellite whereas the CDF=0.99 for the revolving satellite case for 1 kW transmit power.  
\end{itemize}

The remainder of this paper is organized as follows. In Section \ref{System Model}, we explain our dynamic cislunar space model then, the PDF of random harvested power is computed. In Section \ref{Performance Evaluation}, the performances of various EMLP-2 satellite schemes are evaluated based on full SCP and continuous Earth visibility first. The 3D plots showing the contributions of each satellite to the overall SCP of LFS at a sample time are presented. The full SCP rates of various SPS configurations are presented by using our dynamic system model. Then, the CDF of stochastic harvested powers which are extracted from the data are validated by the theoretical CDF obtained in Section \ref{System Model}. Finally, we conclude our paper in Section \ref{Conclusions}.

\vspace{-0.3cm}
\section{SYSTEM MODEL}
\label{System Model}

In our time-based system model, different solar-powered satellite (SPS) configurations which are stable single-satellite on EMLP-2, and revolving single, double, and triple-satellite around the EMLP-2 halo orbit are evaluated based on different semi-major axis length $A_z$ and SCP of the LFS. The distances between the satellites are equal in each multi-satellite scheme. The SPS-1 works actively to transfer laser power to the receiver equipped with solar cells which can be located anywhere on LFS to provide uninterrupted lunar missions in proximity. However, if continuous coverage cannot be satisfied by the first SPS then the second, and even the third satellite can start transferring laser power since a three-satellite scheme with a specific semi-major axis length can provide continuous SCP=100\% for LFS as proved in Section \ref{Performance Evaluation}. 

The harvested power of a free space LoS optical link is expressed by \cite{donmezatp, Shlomi_optimization_2004, ComprehensivePathLoss} as follows
\begin{equation}
{{P}_{H}}\!=\!{{P}_{T}}{{\left( \frac{\lambda }{4\pi R} \right)}^{2}}{{\eta }_{T}}{{\eta}_{H}}{{G}_{T}}{{L}_{T}}({\gamma_{T}}){{G}_{R}}{{L}_{E}}{{L}_{S}}{{L}_{C}},
\label{EQ:1}
\end{equation}
\noindent where ${{P}_{H}}$ is the electrical harvested power, ${{P}_{T}}$ is the electrical (input) transmit power, ${{\eta }_{T}}$ and ${{\eta }_{H}}$ are the wavelength-dependant FSO transmitter PCE and solar cell EHCE, respectively. Moreover, ${{L}_{T}}({\gamma_{T}})$ is the transmitter misalignment loss factor, ${{G}_{T}}$ and ${{G}_{R}}$ are the transmitter and receiver gains, ${{L}_{A}}$ is the atmospheric attenuation loss, ${{L}_{S}}$ is the scintillation loss, and ${{L}_{C}}$ denotes the fiber coupling loss. We consider ${{L}_{E}}={{L}_{S}}={{L}_{C}}=1$ as in \cite{ComprehensivePathLoss}. 

The gains at the transmitting and receiving optical antennas are approximated by, respectively \cite{Shlomi_optimization_2004}. 
{
\begin{align}
  &{{G}_{T}}\approx {{\left( \frac{\pi {{d}_{T}}}{\lambda } \right)}^{2}}, \label{EQ:2a} \\ 
  & {{G}_{R}}\approx {{\left( \frac{\pi {{d}_{R}}}{\lambda } \right)}^{2}},  \label{EQ:2b}
\end{align}
}%
\noindent where ${d}_{T}$ and ${d}_{R}$ are the transmitter and receiver optical antenna diameters, respectively.

The required divergence beam angle of a laser diode can be computed when the spot diameter (i.e., $d_R$) and the LoS distance $d$ are known \cite{spotdiameter} as follows
\begin{equation}
\phi ={{{d}_{R}}}/{d}\;,
\label{EQ:3}
\end{equation}
The corresponding aperture diameter of the transmitter can be found as \cite{OWCmatlab}
\begin{equation}
{{d}_{T}}\cong \lambda /\phi\;,
\label{EQ:4}
\end{equation}
The transmitter misalignment loss can be calculated as \cite{Shlomi_optimization_2004} 
\begin{equation}
{{L}_{T}}({\gamma_{T}})=\exp \left( -{{G}_{T}}\gamma_{T} ^{2} \right),
\label{EQ:5}
\end{equation}
\noindent where ${\gamma_{T}}$ is the radial misalignment error at transmitter.

\begin{figure}[!t]
\centering
\includegraphics[width=\columnwidth]{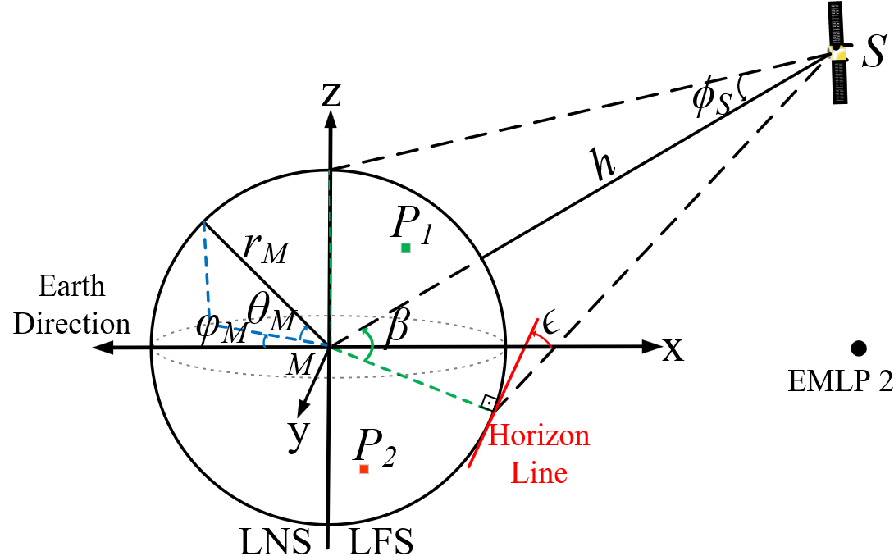}
\caption{Grid point method for lunar coverage analysis.}
\label{fig2}
\vspace{-0.2cm}
\end{figure}

\subsection{Misalignment Error Angle Model}

In modeling the misalignment error for the stable satellite on the EMLP-2, and a satellite revolving around the EMLP-2 halo orbit, the random misalignment induced by the amount of mechanical vibrations is considered in an ascending order. In other words, the maximum possible deviation from the focus point is considered larger for a revolving satellite than the stable one. In addition, the misalignment loss factor ${L}({\gamma})$ is inversely proportional to the LoS distance $R$ between a satellite and the Rx optical antenna. Thus, as the distances are in the order of thousands of kilometers, utilization of the ATP is inevitable to mitigate the pointing errors.

ATP mechanisms that are used for the EMLP-2 satellite and any of the EMLP-2 orbiting identical satellites mitigate the pointing errors in our proposed models and provide high accuracy in the order of nanoscale \cite{nanopointingres}. These errors are considered independent and have different maximum deviation values as presented in Table~\ref{table1}. Monte Carlo simulations are conducted to determine many different misalignment loss factors in (\ref{EQ:5}) and hence the harvested power values can be modeled statistically. 

\begin{table}[!t]
\centering
\caption{Simulation Parameters}
\label{table1}
\begin{tabular}{|l|l|}
\hline
\multicolumn{1}{|c|}{\textbf{Parameter}}                                  & \multicolumn{1}{c|}{\textbf{Value}}                \\ \hline
Transmit Power ($P_T$)                                                     & 1 kW \cite{donmezatp} 
\\ \hline
Laser wavelength ($\lambda$)                                                 & 1064 nm \cite{donmezatp}                               \\ \hline
Laser diode PCE (${{\eta }_{T}}$)                                                      & 51\% \cite{donmezatp}                              \\ \hline
Beam divergence angle (${{\eta }_{T}}$)                                                      & Adaptive  \cite{donmezatp}                              \\ \hline
EHCE (${{\eta}_{H}}$)                                                                 & 50.8\% \cite{Solar1064_PCE51}                           \\ \hline
\begin{tabular}[c]{@{}l@{}}EMLP-2 stable satellite pointing \\ accuracy with ATP (${{\sigma }_{\gamma }}$) \end{tabular} & 5 nrad \cite{lasertypes,nanopointingres}                                            \\ \hline
\begin{tabular}[c]{@{}l@{}}EMLP-2 halo orbit revolving satellite \\ pointing accuracy with ATP (${{\sigma }_{\gamma }}$) \end{tabular}     & 50 nrad \cite{lasertypes,nanopointingres}                  \\ \hline
\begin{tabular}[c]{@{}l@{}}Optical receiver aperture  diameter ($d_r$)\end{tabular} & 1 m \cite{kerslake2008lunar}       \\ \hline
\end{tabular}
\end{table}

The azimuth pointing error angle ${\gamma }_{a}$ and elevation pointing error angle ${\gamma }_{e}$ can be modeled with a Gaussian distribution (i.e., zero-mean) as given \cite{arnon_beam_1997}
{\small
\begin{align}
  &{{f}_{X}}(x)=\frac{1}{\sqrt{2\pi {{\sigma }^{2}}}}\exp \left( -\frac{{{\left( x-\mu  \right)}^{2}}}{2{{\sigma }^{2}}} \right), \nonumber \\ 
  & {{f}_{{{X}_{1}}}}(\left. {{x}_{1}}={{\gamma }_{a}} \right|{{\mu }_{1}}=0,{{\sigma }_{1}}={{\sigma }_{a}})=\frac{1}{\sqrt{2\pi \sigma _{a}^{2}}}\exp \left( -\frac{\gamma _{a}^{2}}{2\sigma _{a}^{2}} \right), \\ \label{EQ:6}
 & {{f}_{{{X}_{2}}}}(\left. {{x}_{2}}={{\gamma }_{e}} \right|{{\mu }_{2}}=0,{{\sigma }_{2}}={{\sigma }_{e}})=\frac{1}{\sqrt{2\pi \sigma _{e}^{2}}}\exp \left( -\frac{\gamma _{e}^{2}}{2\sigma _{e}^{2}} \right),  \nonumber 
\end{align}
}%
\noindent where $\sigma _{a}^{2}$ and $\sigma _{e}^{2}$ are the variances of azimuth and elevation misalignment angles, respectively. These random pointing error angles are independent and identically distributed.

Therefore, the transmitter radial pointing error angle $\gamma$ can be modeled with the Rayleigh distribution by assuming ${{\sigma }_{\gamma }}={{\sigma }_{a}}={{\sigma }_{e}}$ due to the symmetry as expressed~\cite{arnon_beam_1997} 
{
\begin{align}
  &Y=\sqrt{{{X}_{1}}+{{X}_{2}}}, \nonumber \\ 
  & {{f}_{Y}}(y)=\frac{y}{{{\sigma }^{2}}}\exp \left( -\frac{{{y}^{2}}}{2{{\sigma }^{2}}} \right),  \\ \label{EQ:7}
 & {{f}_{Y}}(\left. y=\gamma  \right|\sigma ={{\sigma }_{\gamma }})=\frac{\gamma }{\sigma _{\gamma }^{2}}\exp \left( -\frac{{{\gamma }^{2}}}{2\sigma _{\gamma }^{2}} \right), \nonumber 
\end{align}
}%
Transmitter misalignment loss can be modeled statistically by considering the functions of the random variables approach \cite{leon1994probability,chan2021introduction}. The PDF of the harvested power random variable is determined as follows
{
\begin{equation}
{{f}_{H}}(h)=\frac{{{c}_{2}}}{\left| {{c}_{1}}{{c}_{2}} \right|h}\psi \exp \left( -\psi \left( \frac{\ln \left( {h}/{{{c}_{2}}}\; \right)}{{{c}_{1}}} \right) \right), 
\label{EQ:10} 
\end{equation}
}%
\noindent where ${{c}_{1}}=-{{G}_{T}}$, ${{c}_{2}}={{P}_{T}}{{\left( \frac{\lambda }{4\pi R} \right)}^{2}}{{\eta }_{T}}{{\eta }_{H}}{{G}_{T}}{{G}_{R}}{{L}_{E}}{{L}_{S}}{{L}_{C}}$, and $\psi ={1}/{\left( 2{\sigma _{\gamma }^{2}} \right)}\;$.

\subsection{Surface Coverage Analysis}

The selection of a lunar orbit is vital since none of the aforementioned orbits offers perfect characteristics. Smaller orbits are unfavorable due to the lack of coverage area and instability in general, despite the shorter satellite-to-LFS distances. On the other hand, larger orbits provide sufficient LFS and Earth coverage and stability, however, the maximum LoS distances between a satellite and a receiver optical antenna equipped with solar cells can even exceed 70,000 km for different EMLP-2 halo orbits \cite{L2haloCoverage}. Thus, a higher order of laser power \cite{maximumpower} can be transmitted by a satellite to an Rx optical antenna on LFS.

\begin{table}[!t]
\centering
\caption{Geometric Earth-Moon System Model Parameters}
\label{table-SimPar1}
\begin{tabular}{|l|l|}
\hline
\multicolumn{1}{|c|}{\textbf{Geometric Parameters}}                                            &  \multicolumn{1}{|c|}{\textbf{Values}} \\ \hline
Earth radius                                                                                    & $6,371$ km \cite{MoonEarthFactSheetNASA}                      \\ \hline
Moon radius                                                                                     & $1,737.4$ km \cite{MoonEarthFactSheetNASA}                        \\ \hline
\begin{tabular}[c]{@{}l@{}}Lunar orbit radius \end{tabular} & $385,000$ km \cite{MoonFactsNASA} \\ \hline
\begin{tabular}[c]{@{}l@{}}Earth-Moon Lagrange Point 2 \\ (EMLP-2) distance to Moon\end{tabular}  & $64,500$ km  \cite{farquhar1971utilization}                        \\ \hline
Earth obliquity                                                                                 & $23.44^\circ$   \cite{olthoff2023dynamic}                    \\ \hline
Lunar obliquity                                                                                 & $6.68^\circ$       \cite{olthoff2023dynamic}                  \\ \hline
Lunar obliquity to ecliptic plane                                                               & $1.54^\circ$      \cite{olthoff2023dynamic}                   \\ \hline
\begin{tabular}[c]{@{}l@{}}Lunar orbital plane inclination\\ to Ecliptic plane\end{tabular}     & $5.14^\circ$    \cite{olthoff2023dynamic}                     \\ \hline
\end{tabular}%
\end{table}

Our goal is to provide continuous energy harvesting from the SPS to an easily transportable circular receiver (i.e., limited $d_R$), which can even be located on the curve where LFS and lunar near side (LNS) coincide. Therefore, different satellite schemes and orbit sizes \cite{farquhar1971utilization,orbittypes} are evaluated to determine a benchmark model that offers full coverage (i.e., SCP=100\%) for the LFS. The semi-minor axis length $A_y$ is chosen larger than 3671 km to provide continuous Earth visibility and has a relation with $A_y=0.343\times A_z$ to reduce the energy requirement to remain in the halo orbit \cite{farquhar1967,farquhar1971utilization}. This scheme is considered as a benchmark and the LoS distances of links are used for modeling the harvested power statistically.

The uniform sampling method is used to determine the grid points on the lunar surface by considering the sampling interval for latitude angle, $\theta_M$, and longitude angle, $\varphi_M$, as $\Delta {{\theta }_{M}}=\Delta {{\varphi }_{M}}={{1}^{\circ }}$. The grid point approach \cite{cakaj2023CentANG,song2018novel,gao2020SCP} is applied to find the surface coverage percentage (SCP) of the lunar far side which has a latitude and longitude ranges of (-90$^\circ$,90$^\circ$). However, the latitude range for the LFS southern pole is considered as (-90$^\circ$, -80$^\circ$) \cite{LunarSouthPole}. 

The largest possible lunar surface coverage can be achieved when the elevation angle $\epsilon$, between the satellite and the horizon line, becomes zero. Since we assume that there is no constraint on the moveable laser transmitter of the satellites, the largest possible surface coverages can be achieved in each instant of our dynamic system model. 

When the largest possible coverage is attained, the nadir angle $\phi_S$ takes its maximum value. The nadir angle $\phi_S$ can be determined by the elevation angle $\epsilon$ as follows~\cite{cakaj2023CentANG}  
\begin{equation}
\sin \phi_S =\frac{{{r}_{M}}}{{{r}_{M}}+h}\cos \epsilon,
\label{EQ:11}
\end{equation}
\noindent where $r_M$ denotes the moon radius which is 1737.4 km, $h$ is the altitude. 

The elevation angle is considered as $\epsilon=0$ and hence the nadir angle $\phi_S$ can be rewritten as follows
\begin{equation}
\phi_S =\arcsin \left( \frac{{{r}_{M}}}{{{r}_{M}}+h} \right),
\label{EQ:12}
\end{equation}
Since the elevation angle $\epsilon$, nadir angle $\phi_S$, and central angle $\beta$ equals to $90^\circ$ as shown in Fig. \ref{fig2}, and $\epsilon$ is also considered zero therefore $\beta=90^\circ-\phi_S$.  

The coverage condition of a grid point, $P$, on the lunar far side surface from a satellite located at point $S$ in the EMLP-2 orbit is given as \cite{song2018novel} 
\begin{equation}
\arccos \left( \frac{\overrightarrow{MP}\,\cdot \,\overrightarrow{MS}}{\left\| \overrightarrow{MP} \right\|\left\| \overrightarrow{MS} \right\|} \right)<\beta,
\label{EQ:13}
\end{equation}
\noindent where $M$ is the lunar centroid thus, $\overrightarrow{MP}$ and $\overrightarrow{MS}$ are the vectors that represent the displacements between $M$ and  $P$, and between $M$ and $S$, respectively. As per Fig. \ref{fig2}, $P_1$ and $P_2$ are illustrated as the covered and uncovered points, respectively.

The surface coverage percentage can be computed precisely by considering each grid area as ${r_{M}^{2}}\cos {{\theta }_{M}}\Delta {{\theta }_{M}}\Delta {{\varphi }_{M}}$, and SCP can be computed by \cite{gao2020SCP} 
\begin{equation}
SCP=\frac{\sum\nolimits_{i\in S}{{{\zeta }_{i}}\,r_{M}^{2}\cos {{\theta }_{M}}\Delta {{\theta }_{M}}\Delta {{\varphi }_{M}}}}{\sum\nolimits_{i\in S}{r_{M}^{2}\cos {{\theta }_{M}}\Delta {{\theta }_{M}}\Delta {{\varphi }_{M}}}},
\label{EQ:14}
\end{equation}
\noindent where ${{\zeta }_{i}}$ is a function that takes the value of 1 when $i^{th}$ grid point is covered, elsewhere takes zero.

\begin{table}[!t] 
\centering
\caption{Temporal Earth-Moon System Model Parameters}
\label{table-SimPar2}
\begin{tabular}{|l|l|}
\hline
\multicolumn{1}{|c|}{\textbf{Temporal Parameters}} & \multicolumn{1}{|c|}{\textbf{Values}} \\ \hline
Earth's rotation              & 1 day \cite{EarthInfoNASA}          \\ \hline
Moon's rotation               & 27 days \cite{MoonFactsNASA}        \\ \hline
Moon's revolution             & 27 days \cite{MoonFactsNASA}        \\ \hline
\begin{tabular}[c]{@{}l@{}}Earth-Moon Lagrange Point 2 \\ (EMLP-2) halo orbit period\end{tabular} & 8 days \cite{orbittypes}    \\ \hline
Total simulation time                                                                                    &  $648$ hours                   \\ \hline
Sampling time                                                                                    & $1$ hour                     \\ \hline
Angular velocity of Earth                                                                                    &  $15^\circ$/hour \cite{EarthInfoNASA}      \\ \hline
Angular velocity of Moon                                                                                    & $0.56^\circ$/hour   \cite{MoonFactsNASA}     \\ \hline
Laser propagation time                                                                                   & $\approx 0.2$ sec        \\ \hline
\end{tabular}%
\end{table}

\section{PERFORMANCE EVALUATION}
\label{Performance Evaluation}

The SCP performances of different satellite configurations and orbit sizes are evaluated to determine a benchmark that offers continuous LFS SCP=100\% and Earth visibility. Then we compare the full coverage duration per period performances of these satellite configurations for a specific region such as the lunar south pole where ongoing researches are conducted. Since the misalignment errors are random, we model the harvested power by the Rx optical antenna equipped with solar cells statistically, based on the varying distances of the orbiting satellites in our benchmark model. 

\subsection{Simulation Settings}
\label{Simulation Parameters}

The simulation parameters are presented in Table \ref{table1}. The laser diode with 1064 nm wavelength is selected as in \cite{donmezatp} since it is suitable for energy harvesting and deep-space missions. The average PCE is considered as 51\% for the 1064 nm laser source \cite{laser1064}. As EHCE needs to be consistent with the laser source wavelength, InGaAs-made solar cells which offer 50.8$\%$ are selected in our system model~\cite{Solar1064_PCE51}. The circular Rx optical antenna diameter is considered as 1 m which facilitates its installation and transportation \cite{kerslake2008lunar}. It is a challenging task to achieve a very narrow beam divergence angle to obtain a smaller spot diameter at far distances in the order of thousands of kilometers since larger transmitter aperture diameters are required such that $d_T=$ 50 m is computed in \cite{Txaperturedia}. 

The dynamic system model of the benchmark scheme which consists of the equidistant three satellites is presented in Fig. \ref{fig3}. Both geometric and temporal Earth-Moon system model parameters shown in Tables \ref{table-SimPar1} and \ref{table-SimPar2} are considered in our simulations. In addition, Earth visibility can be attained with the closest deep space network station (DSNS) \cite{DSNlocations2022} rotating with the Earth-centered Earth-fixed (ECEF) reference frame \cite{farrell2008aided} as well. The affine transformations in which intrinsic rotation and translation transforms \cite{rodrigues1840lois,paul1981robot,house2016foundations} are utilized to determine the time-varying locations of the system elements at every new hour until 648 hours, or one revolution of the Moon around the Earth. This dynamic system allows us to determine a specific region's full coverage (i.e., SCP=100\%) duration within a period.

\subsection{Results and Discussions}
\label{Results and Discussions}

\begin{figure*}[ht]
\centering
\includegraphics[width=0.8\textwidth]{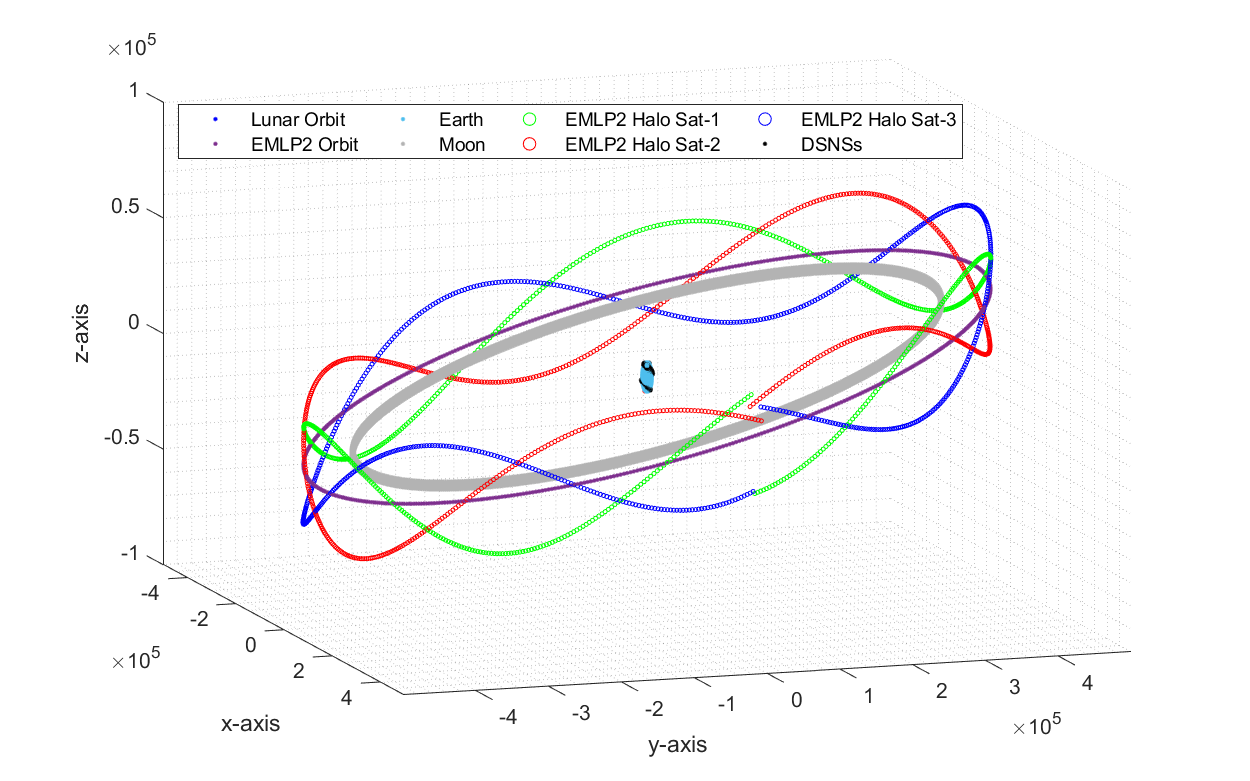}
\caption{Time-based triple satellite Earth-Moon system model}
\label{fig3}
\vspace{-0.2cm}
\end{figure*}

\begin{figure*}[!ht]
\centering
\subfloat[]{
	\label{subfig:4a}
	\includegraphics[clip, scale=0.5]{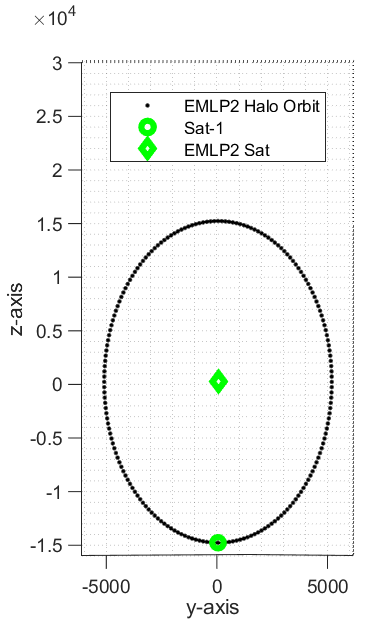}
	 }
\subfloat[]{
	\label{subfig:4b}
	\includegraphics[clip, scale=0.5]{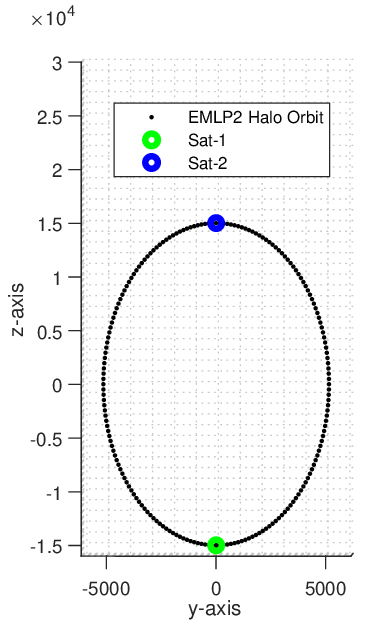}
	 }
\subfloat[]{
	\label{subfig:4c}
	\includegraphics[clip, scale=0.5]{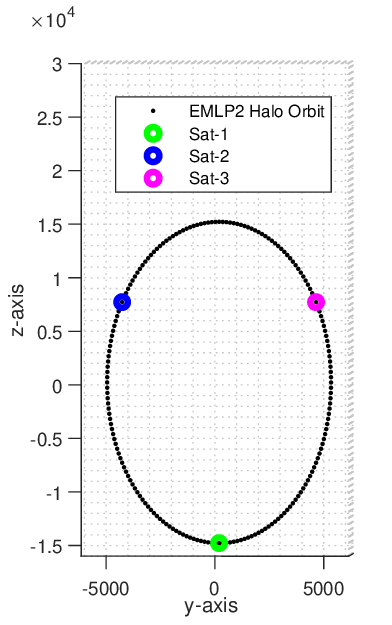}
	 }

\caption{EMLP-2 satellite schemes (a) Single satellite, (b) Double satellite, (c) Triple satellite}
\label{fig4}
\vspace{-0.15cm}
\end{figure*}

\begin{figure}[t!]
\centering
\includegraphics[width=.45\textwidth]{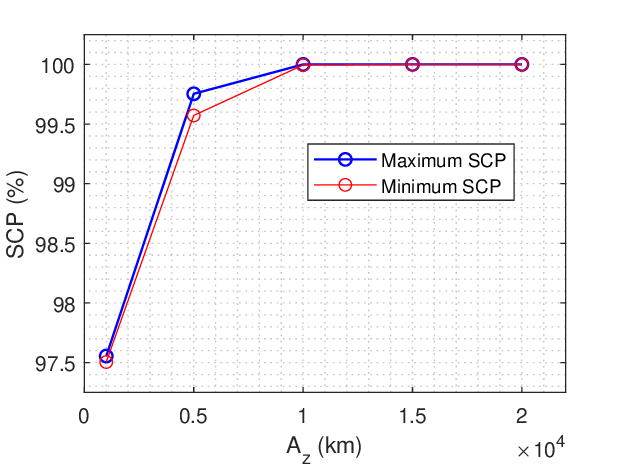}
\caption{LFS SCP values for various $A_z$ in the triple satellite scheme}
\label{fig5}
\vspace{-0.2cm}
\end{figure}

\begin{figure}[t!]
\centering
\includegraphics[width=.45\textwidth]{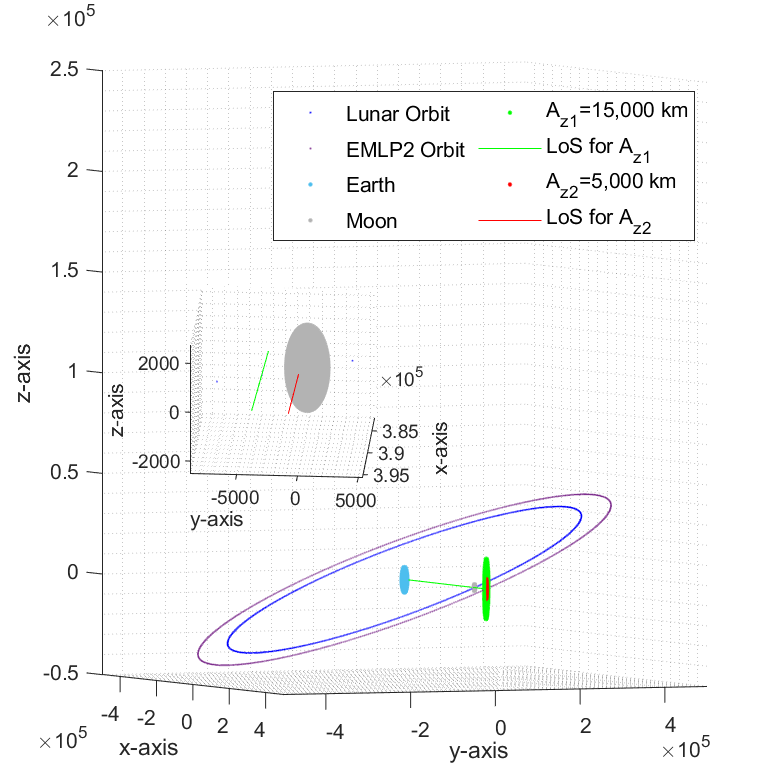}
\caption{Earth visibility from the EMLP-2 halo orbit with different semi-major axis lengths $A_z$ }
\label{fig6}
\vspace{-0.2cm}
\end{figure}

Various satellite schemes and halo orbit sizes are evaluated first to define a benchmark that offers continuous LFS SCP=100\% and Earth visibility (i.e., DSNS) in our dynamic cislunar space model presented in Fig. \ref{fig3}. The overall SCP values for different satellite scenarios whose initial positions are presented in Fig. \ref{fig4} are obtained at each sampling time during the orbital period of 192 hours (Table \ref{table-SimPar2}). Single and double satellite scenarios cannot achieve SCP=100\% for different $A_z$ values \cite{orbittypes} but the three-satellite configuration does. However, not all three satellite schemes can guarantee continuous full LFS SCP as is presented in Fig. \ref{fig5} thus, different $A_z$ values are evaluated as well. For instance, halo orbit with $A_z=5\times10^3$ is not only unsuccessful in providing SCP=100\% but also fails in enabling continuous Earth visibility as demonstrated in Fig. \ref{fig6}. Therefore, a three-satellite scheme moving around the halo orbit with $A_z=15\times10^3$ km is accepted as our benchmark to compare with others since it becomes successful in terms of continuous full SCP of LFS and Earth visibility as proved in both Figs. \ref{fig5} and \ref{fig6}.  

\begin{figure*}[!ht]
\centering

\subfloat[]{
	\label{subfig:5a}
	\includegraphics[clip, scale=0.62]{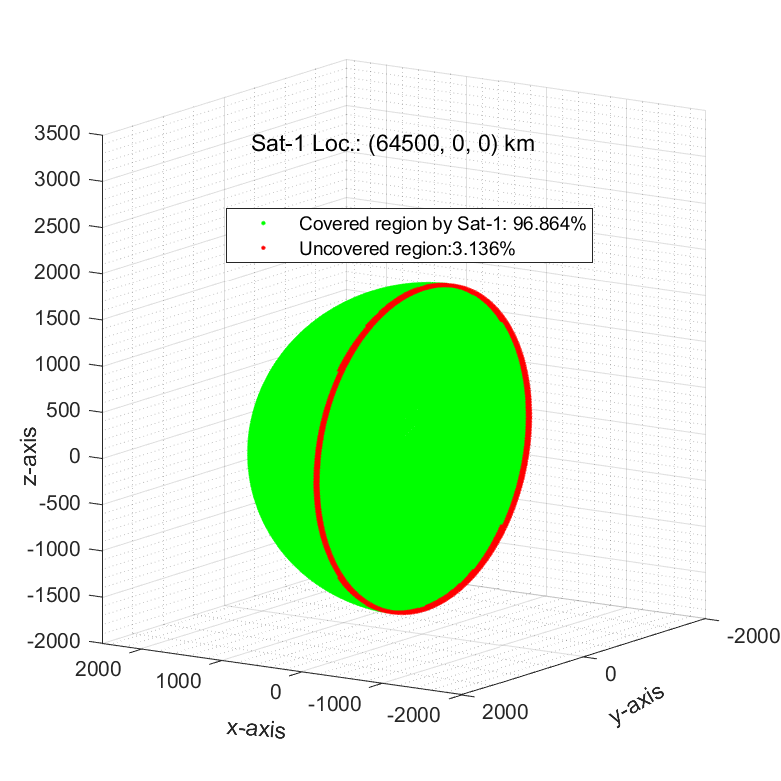}
	 }
\subfloat[]{
	\label{subfig:5b}
	\includegraphics[clip, scale=0.62]{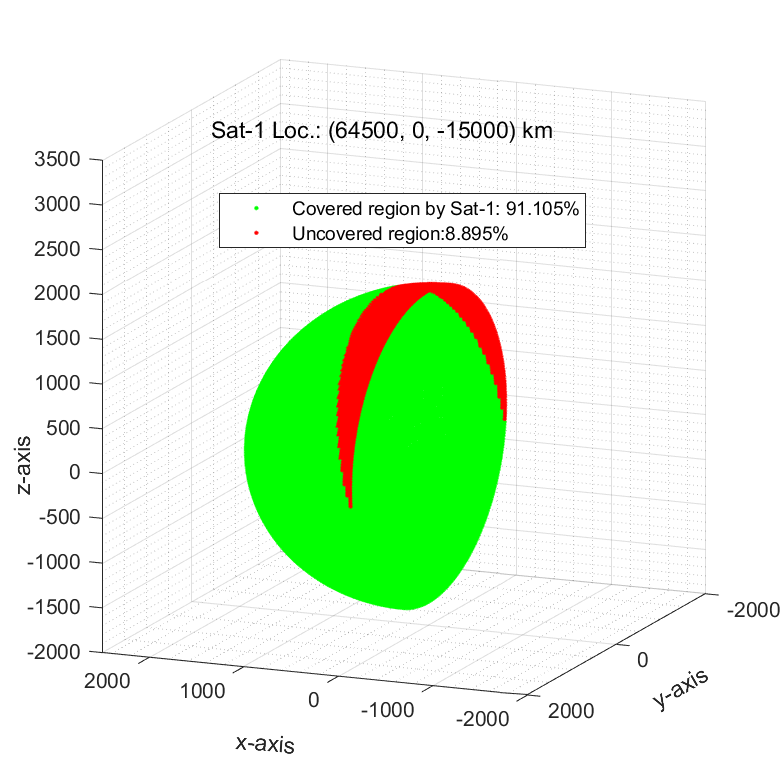}
	 }

\subfloat[]{
	\label{subfig:5c}
	\includegraphics[clip, scale=0.62]{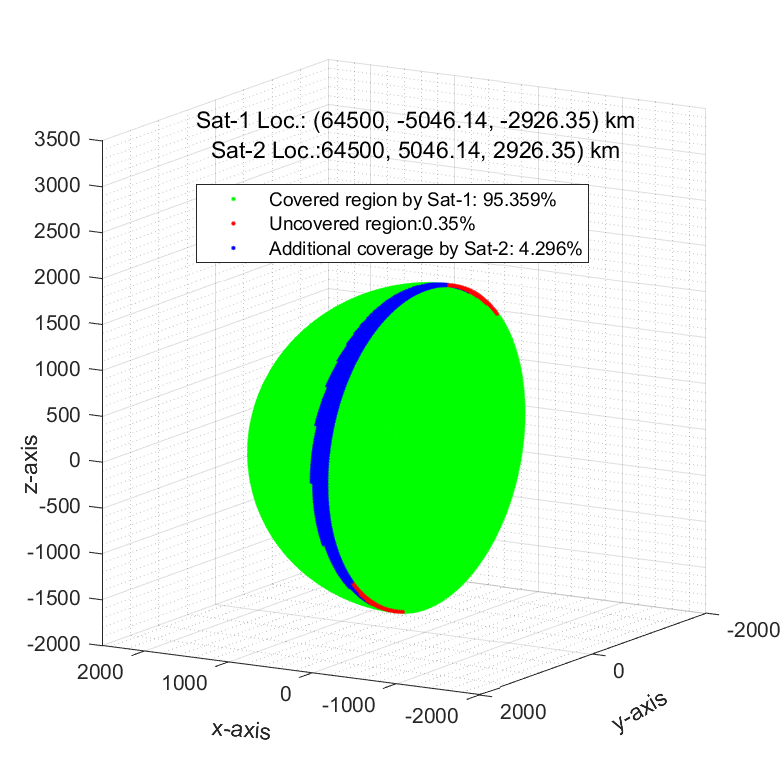}
	 }
\subfloat[]{
	\label{subfig:5d}
	\includegraphics[clip, scale=0.62]{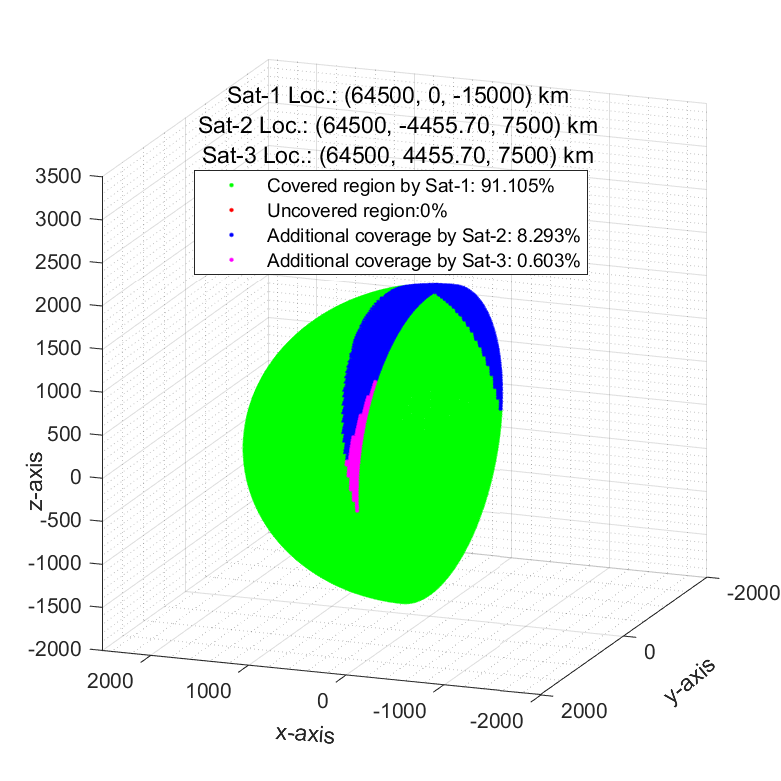}
	 }
\\
\caption{Minimum surface coverage (SCP) scenarios based on lunar reference frame for different satellite schemes  (a) stable single satellite, (b) revolving single satellite, (c) revolving two satellites, (d) revolving three satellites }
\label{fig7}
\vspace{-0.15cm}
\end{figure*}

To elaborate on the SCP impact of each satellite which shows an effort to minimize the uncovered LFS regions that remain from the preceding satellite, the 3D LFS figures are presented in Fig. \ref{fig7}. It can be inferred that the satellite on EMLP-2 is not successful in power beaming to the contours of LFS although it provides SCP=96.864\% whereas in orbiting single satellite scenario, each and every point of the LFS can be covered but for a while only. In multi-satellite scenarios, the second satellite shows an effort to attain overall SCP=100\% whereas our benchmark model achieves an overall full LFS SCP throughout a period. To compare the SCP performances of these configurations based on the time, the ratio of the duration for full coverage of a specific region such as the promising southern pole \cite{LunarSouthPole}, over the satellite orbital period of EMLP-2 is considered. It is found that an equidistant 2-satellite scheme provides SCP=100\% at 88.60\% of the orbital period of 8 days.

\begin{figure*}[!h]
\centering

\subfloat[]{
	\label{subfig:8a}
	\includegraphics[clip, scale=.85]{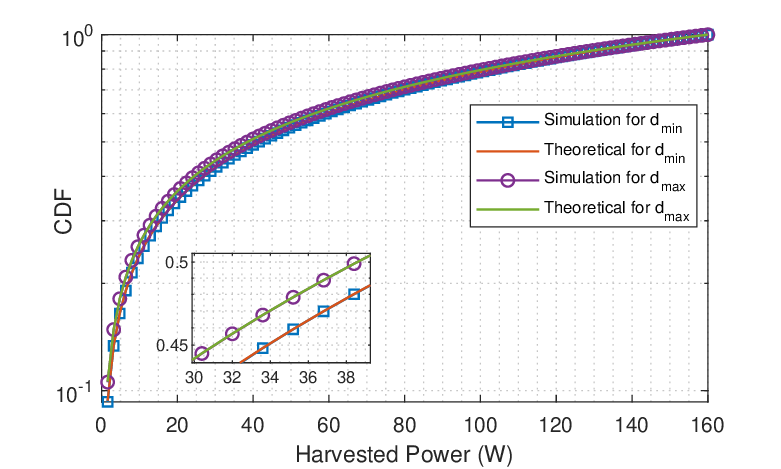}
	 }
  \\
\subfloat[]{
	\label{subfig:8b}
	\includegraphics[clip, scale=.85]{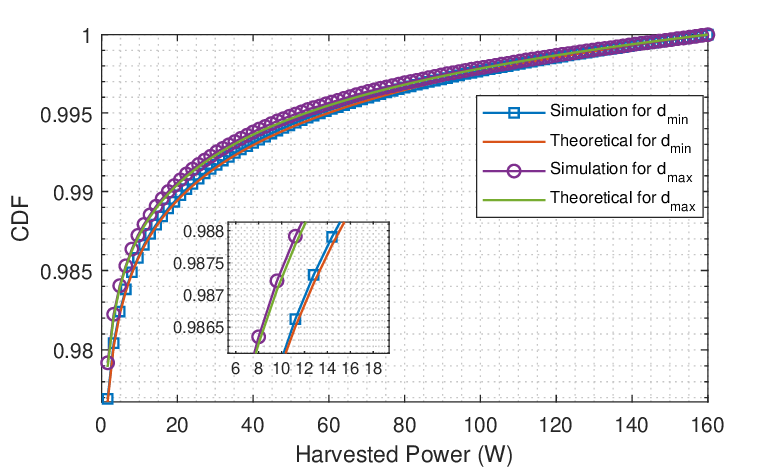}
	 }
\caption{Statistical models of harvested power $P_H$ for (a) stable EMLP-2 satellite, and (b) revolving EMLP-2 halo satellite}
\label{fig8}
\vspace{-0.15cm}
\end{figure*}

The pointing accuracy of a stable EMLP-2 satellite is better than that of a satellite revolving around EMLP-2 as considered in Table \ref{table1}. In other words, the statistical model of the random misalignment radial angle spreads more for a revolving satellite. The minimum and maximum covered distances are computed along with the surface coverage analyses first for the former and the latter schemes. 
The CDFs of harvested powers extracted from the simulations are validated with the theoretical results as presented in Fig. \ref{fig8}. The probability of $P_H\le$1.6~W is around 0.1 for the stable satellite whereas the CDF=0.98 for the revolving satellite case for 1 kW transmit power. It can be inferred that the harvested power is expected to be around $P_H\le$1.6W in revolving satellite scenarios whereas much higher power can be generated from a stable satellite but not on the contour of LFS.   
      
\begin{table}[!t]
\centering
\caption{Time rate comparison for LFS south pole SCP=$100\%$}
\label{table-LFS_SP_SCP_Duration}
\begin{tabular}{|ccc|}
\hline
\multicolumn{3}{|c|}{LFS SP Full SCP Duration}                                     \\ \hline
\multicolumn{1}{|c|}{1-Satellite} & \multicolumn{1}{c|}{2-Satellite} & 3-Satellite \\ \hline
\multicolumn{1}{|c|}{44.56\%}         & \multicolumn{1}{c|}{88.60\%}         & 100\%         \\ \hline
\end{tabular}%
\end{table}

\section{CONCLUSIONS}
\label{Conclusions}

In this paper, we evaluated different satellite configurations and EMLP-2 orbit sizes that provide continuous full lunar far-side surface coverage and Earth visibility while adhering to the semi-major and -minor axis length ratio that minimizes the stationkeeping cost. Single and equidistant multiple satellite schemes in Fig. \ref{fig4} were investigated with different semi-different major axis lengths to define a benchmark with a minimum number of satellites. We found that a three-satellite scenario with a semi-major axis length of 15,000 km attains continuous LFS full coverage and Earth visibility as proved in Fig. \ref{fig5} and \ref{fig6}, respectively.

Our dynamic cislunar space model presented in Fig. \ref{fig3} was generated by considering the geometric and temporal features of the Earth-Moon system. This model enables us to investigate the time-based performances of different scenarios having the same EMLP-2 orbit as the benchmark scheme. It was found that single and double satellite schemes provide full coverage of the LFS southern pole at 44.56\% and 88.60\% of the orbital period of 192 hours (Table \ref{table-LFS_SP_SCP_Duration}).

The impact of each satellite on overall SCP is elaborated in Fig. \ref{fig7} and it is found that a stable single satellite cannot cover the edges where LFS and LNS coincide however a revolving satellite on the halo orbit can power beam a receiver optical antenna equipped with solar cells anywhere on LFS if not all the time. The two-satellite scenario shows that equidistant satellites almost achieve a full coverage whereas our benchmark scenario with three satellites becomes successful with a little contribution of the third one.  

 Finally, the stochastic harvested powers extracted from the stable EMLP-2 satellite and a revolving halo orbit satellite scenarios in our simulations are validated by the theoretical cumulative distribution function computed from the previously determined probability density function (Figs. \ref{fig8}). The probability of $P_H\le$1.6~W is around 0.1 for the stable satellite whereas the CDF=0.98 for the revolving satellite case for 1 kW transmit power.

\bibliographystyle{IEEEtaes}
%

%
\begin{IEEEbiography}
[{\includegraphics[width=1in,height=1.25in,clip,keepaspectratio]{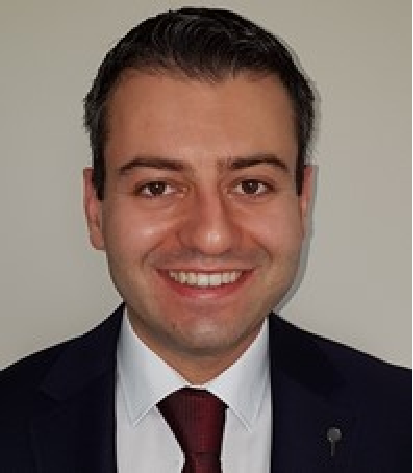}}] {Baris Donmez}{\space}(Graduate Student Member, IEEE) received the B.Sc. and M.Sc. degrees (with high honors) in electrical and electronics engineering, in 2009 and 2022, respectively, from FMV Işık University, Istanbul, Turkiye. Now, he is currently working toward a Ph.D. degree in electrical engineering at Polytechnique Montréal, Montréal, QC, Canada

His research interests include communication and energy harvesting systems in deep-space networks.
\end{IEEEbiography}

\begin{IEEEbiography}
[{\includegraphics[width=1in,height=1.25in,clip,keepaspectratio]{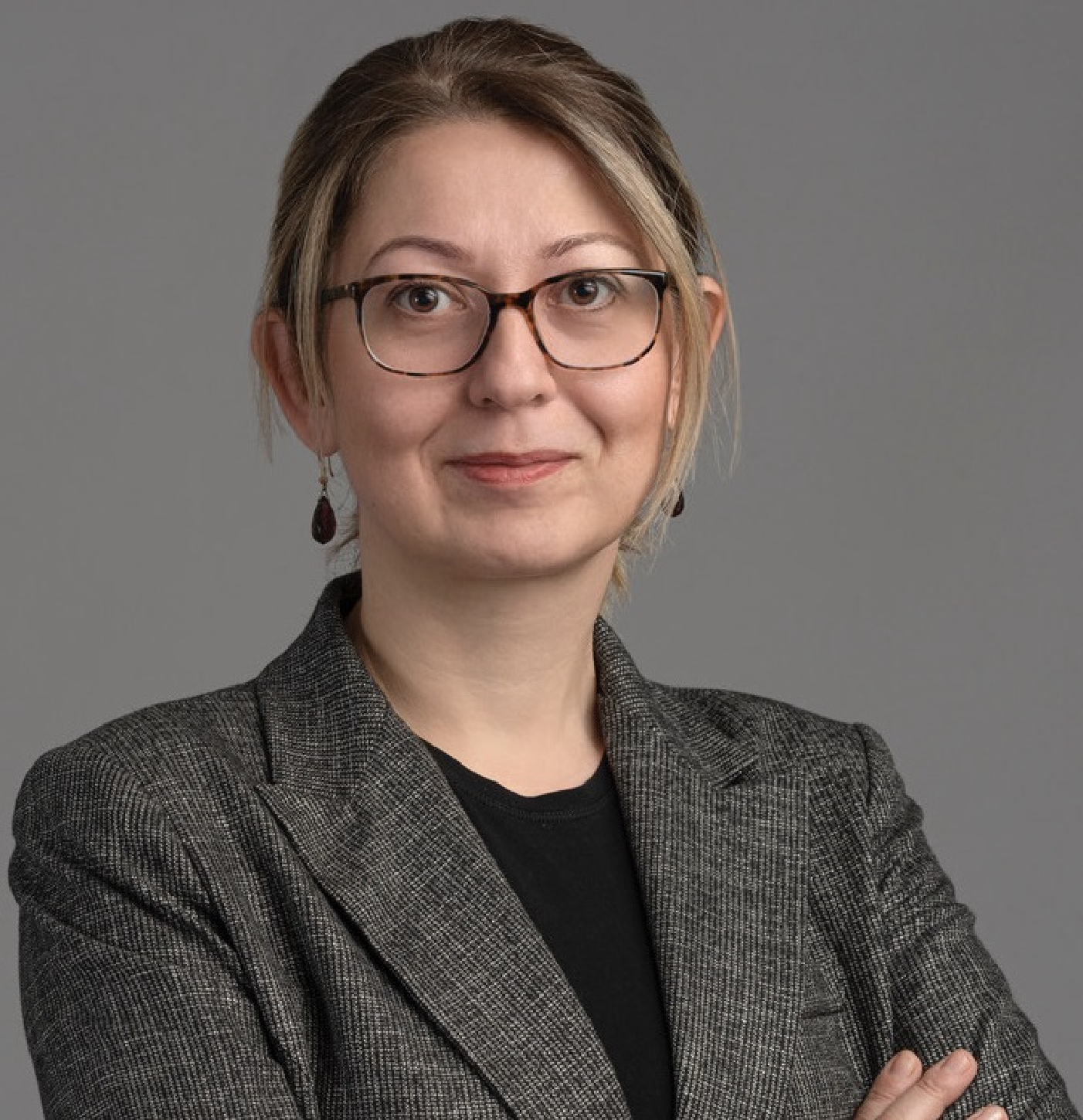}}] {Gunes Karabulut Kurt}{\space}(Senior Member, IEEE) is a Canada Research Chair (Tier 1) in New Frontiers in Space Communications and Associate Professor at Polytechnique Montréal, Montréal, QC, Canada. She is also an adjunct research professor at Carleton University, ON, Canada. Gunes is a Marie Curie Fellow and has received the Turkish Academy of Sciences Outstanding Young Scientist (TÜBA-GEBIP) Award in 2019. She received her Ph.D. degree in electrical engineering from the University of Ottawa, ON, Canada. 

\end{IEEEbiography}

\end{document}